\newcommand{\gtrsim}{\,\rlap{\lower3.5pt\hbox{$\mathchar\sim$}}
\raise1pt\hbox{$>$}\,}
\newcommand{\lesssim}{\,\rlap{\lower3.5pt\hbox{$\mathchar\sim$}}
\raise1pt\hbox{$<$}\,}

\documentclass{ar2e}
\usepackage{citesort}
\usepackage{epsfig}

\begin{document}
\input epsf.tex    %<-If you need EPS figures to be
                   %  called in {figure} environment for PC
%\input epsf.def   %<-If you need EPS figures to be
                   %  called in {figure} environment for Macintosh

\input psfig.sty

\jname{Annu. Rev. Nucl. Part. Phys.} \jyear{2006} \jvol{1}
\ARinfo{1056-8700/97/0610-00}

\title{Primordial Neutrinos}

\author{Steen Hannestad
\affiliation{Department of Physics and Astronomy, University of
Aarhus, Ny Munkegade, DK-8000 Aarhus C, Denmark}}


\begin{abstract}
The connection between cosmological observations and neutrino
physics is discussed in detail. Neutrinos decouple from thermal
contact in the early Universe at a temperature of order 1 MeV
which coincides with the temperature where light element synthesis
occurs. Observation of light element abundances therefore provides
important information on such properties as neutrino energy
density and chemical potential. Precision observations of the
cosmic microwave background and large scale structure of galaxies
can be used to probe neutrino masses with greater precision than
current laboratory experiments. In this review I discuss current
cosmological bounds on neutrino properties, as well as possible
bounds from upcoming measurements.
\end{abstract}

\maketitle

%%%%%%%%%%%%%%%%%%%%%%%%%%%%%%%%%%%%%%%%%%%%%%%%%%%%%%%%%%%%%%%%%%%%%%
\section{Introduction} %%%%%%%%%%%%%%%%%%%%%%%%%%%%%%%%%%%%%%%%%%%%%%%
%%%%%%%%%%%%%%%%%%%%%%%%%%%%%%%%%%%%%%%%%%%%%%%%%%%%%%%%%%%%%%%%%%%%%%

In the past few years a new standard model of cosmology has been
established in which most of the energy density of the Universe is
made up of cold dark matter and a component with negative
pressure, generically referred to as dark energy. This model
provides an amazingly good fit to all observational data with
relatively few free parameters and has allowed for stringent
constraints on the basic cosmological parameters.

The precision of the data is now at a level where observations of
the cosmic microwave background (CMB), the large scale structure
(LSS) of galaxies, and type Ia supernovae can be used to probe
important aspects of particle physics such as neutrino properties.
Conversely, cosmology is now also at a level where unknowns from
the particle physics side can significantly bias estimates of
cosmological parameters.

The prime example of this interplay between particle physics and
cosmology is the use of cosmological data to probe neutrino
physics. Particularly the possibility to constrain the neutrino
mass using cosmological measurements has been studied extensively.

If only the three known neutrino flavour states, $\nu_f$, are
considered, they must correspond to three mass eigenstates,
$\nu_i$. The two sets are related by
\begin{equation}
\nu_f = U \nu_i,
\end{equation}
where $U$ is the $3 \times 3$ neutrino mixing matrix.

The neutrino oscillation probability in the general $3 \times 3$
oscillation case is quite complicated, but always depends on
$\delta m^2$, i.e.\ on squared mass differences. The combination
of all currently available data from neutrino oscillation
experiments suggests two important mass differences in the
neutrino mass hierarchy. The solar mass difference of $\Delta
m_{12}^2 \simeq 7.5 - 8.7 \times 10^{-5}$ eV$^2$ $(2\sigma)$ and
the atmospheric mass difference $|\Delta m_{23}^2| \simeq 1.7 -
2.9 \times 10^{-3}$ eV$^2$ $(2\sigma)$ \cite{Maltoni:2004ei}. In
the simplest case where neutrino masses are hierarchical these
results suggest that $m_1 \sim 0$, $m_2 \sim \Delta m_{\rm
solar}$, and $m_3 \sim \Delta m_{\rm atmospheric}$. If the
hierarchy is inverted one instead finds $m_3 \sim 0$, $m_2 \sim
\Delta m_{\rm atmospheric}$, and $m_1 \sim \Delta m_{\rm
atmospheric}$. However, it is also possible that neutrino masses
are degenerate, $m_1 \sim m_2 \sim m_3 \gg \Delta m_{\rm
atmospheric}$. Since oscillation probabilities depend only on
squared mass differences, $\Delta m^2$, such experiments have no
sensitivity to the absolute value of neutrino masses, and if the
masses are degenerate oscillation experiments are not useful for
determining the absolute mass scale.

Experiments which rely on kinematical effects of the neutrino mass
offer the most robust probe of this overall mass scale. Tritium
decay measurements by the Mainz experiment have been able to put
an upper limit on the effective electron neutrino mass of
$m_{\nu_e} = \left( \sum_i  |U_{ei}|^2 m_i^2\right)^{1/2}  < 2.3$
eV (95\% conf.) \cite{kraus} (note that this is an incoherent sum
so that cancellation due to phases cannot occur). However,
cosmology at present provides an even better limit, as will be
discussed in detail in the next sections.

Very interestingly there is also a claim of direct detection of
neutrinoless double beta decay in the Heidelberg-Moscow experiment
\cite{Klapdor-Kleingrothaus:2001ke,Klapdor-Kleingrothaus:2004wj,%
VolkerKlapdor-Kleingrothaus:2005qv}. Neutrinoless double beta
decay is only possible if neutrinos are majorana particles because
it requires violation of lepton number. The lifetime of the
neutrinoless mode of the decay is inversely proportional to the
square of the effective mass of the neutrino because it involves
flipping the spin of the internal neutrino.

The claimed lifetime of $^{76}Ge$ measured by the experiment can
be translated to a neutrino mass. However, there is a large
uncertainty in this translation because the involved nuclear
matrix element is quite difficult to calculate (see for instance
\cite{Faessler:2005zj}). Including the estimated matrix element
uncertainty the result corresponds to an effective neutrino mass
in the $0.1-0.9$ eV range for the parameter $m_{ee} = \left|
\sum_j U^2_{ej} m_{\nu_j} \right|$ (note that, contrary to the
tritium measurement, this is a coherent sum which means that it
can be suppressed by phases in the mixing matrix). If this result
is confirmed then it shows that neutrino masses are almost
degenerate.

Another important question which can be answered by cosmological
observations is how large the total neutrino energy density is.
Apart from the standard model prediction of three light neutrinos,
such energy density can be either in the form of additional,
sterile neutrino degrees of freedom, or a non-zero neutrino
chemical potential.

In this review I discuss mainly these two questions. Due to the
limited space available other interesting aspects of neutrino
cosmology, such as leptogenesis, have been left out. The
interested reader is referred to the very thorough review
\cite{dolgov}.

The paper is divided into sections in the following way: In
section 2 I review the thermal evolution of light neutrinos,
particularly aspects related to the decoupling of neutrinos around
a temperature of 1 MeV. I also discuss Big Bang nucleosynthesis
and its relation to neutrino physics.

I section 3 I discuss the current cosmological data available, and
section 4 contains a review of bounds on neutrino properties from
this data. Finally, section 5 contains a discussion.

%%%%%%%%%%%%%%%%%%%%%%%%%%%%%%%%%%%%%%%%%%%%%%%%%%%%%%%%%%%%%%%%%%%%%%
\section{The thermal history of light neutrinos} %%%%%%%%%%%%%%%%%%%%%
%%%%%%%%%%%%%%%%%%%%%%%%%%%%%%%%%%%%%%%%%%%%%%%%%%%%%%%%%%%%%%%%%%%%%%

\subsection{Standard model}

In the standard model neutrinos interact via weak interactions
with charged leptons, keeping them in equilibrium with the
electromagnetic plasma at high temperatures. Below $T \sim 30-40$
MeV $e^+$ and $e^-$ are the only relevant particles, greatly
reducing the number of possible reactions which must be
considered. In the absence of oscillations neutrino decoupling can
be followed via the Boltzmann equation for the single particle
distribution function \cite{kolb}
\begin{equation}
\frac{\partial f}{\partial t} - H p \frac{\partial f}{\partial p}
= C_{\rm coll}, \label{eq:boltz}
\end{equation}
where $C_{\rm coll}$ represents all elastic and inelastic
interactions. In the standard model all these interactions are $2
\leftrightarrow 2$ interactions in which case the collision
integral for process $i$ can be written
\begin{eqnarray}
C_{\rm coll,i} (f_1) & = & \frac{1}{2E_1} \int \frac{d^3 {\bf
p}_2}{2E_2 (2\pi)^3} \frac{d^3 {\bf p}_3}{2E_3 (2\pi)^3} \frac{d^3
{\bf p}_4}{2E_4 (2\pi)^3} \nonumber \\
&& \,\, \times (2\pi)^4 \delta^4
(p_1+p_2-p_3+p_4)\Lambda(f_1,f_2,f_3,f_4) S |M|^2_{12 \to 34,i},
\end{eqnarray}
where $S |M|^2_{12 \to 34,i}$ is the spin-summed and averaged
matrix element including the symmetry factor $S=1/2$ if there are
identical particles in initial or final states. The phase-space
factor is $\Lambda(f_1,f_2,f_3,f_4) = f_3 f_4 (1-f_1)(1-f_2) - f_1
f_2 (1-f_3)(1-f_4)$.

The matrix elements for all relevant processes can for instance be
found in Ref.~\cite{Hannestad:1995rs}. If Maxwell-Boltzmann
statistics is used for all particles, and neutrinos are assumed to
be in complete scattering equilbrium so that they can be
represented by a single temperature, then the collision integral
can be integrated to yield the average annihilation rate for a
neutrino
\begin{equation}
\Gamma = \frac{16 G_F^2}{\pi^3} (g_L^2 + g_R^2) T^5,
\end{equation}
where
\begin{equation}
g_L^2 + g_R^2 = \cases{\sin^4 \theta_W + (\frac{1}{2}+\sin^2
\theta_W)^2 & for $\nu_e$ \cr \sin^4 \theta_W +
(-\frac{1}{2}+\sin^2 \theta_W)^2 & for $\nu_{\mu,\tau}$}.
\end{equation}

This rate can then be compared with the Hubble expansion rate
\begin{equation}
H = 1.66 g_*^{1/2} \frac{T^2}{M_{\rm Pl}} \label{eq:hub}
\end{equation}

 to
find the decoupling temperature from the criterion $\left. H =
\Gamma \right|_{T=T_D}$. From this one finds that $T_D(\nu_e)
\simeq 2.4$ MeV, $T_D(\nu_{\mu,\tau}) \simeq 3.7$ MeV, when $g_*
=10.75$, as is the case in the standard model.

This means that neutrinos decouple at a temperature which is
significantly higher than the electron mass. When $e^+e^-$
annihilation occurs around $T \sim m_e/3$, the neutrino
temperature is unaffected whereas the photon temperature is heated
by a factor $(11/4)^{1/3}$. The relation $T_\nu/T_\gamma =
(4/11)^{1/3} \simeq 0.71$ holds to a precision of roughly one
percent. The main correction comes from a slight heating of
neutrinos by $e^+e^-$ annihilation, as well as finite temperature
QED effects on the photon propagator
\cite{Dicus:1982bz,Rana:1991xk,herrera,Dolgov:1992qg,Dodelson:1992km,%
Fields:1993zb,Hannestad:1995rs,Dolgov:1997mb,Dolgov:1999sf,gnedin,%
Esposito:2000hi,Steigman:2001px,Mangano:2001iu,osc,Mangano:2005cc}.

\subsection{Big Bang nucleosynthesis and the number of neutrino species}

Shortly after neutrino decoupling the weak interactions which keep
neutrons and protons in statistical equilibrium freeze out. Again
the criterion $\left. H = \Gamma \right|_{T=T_{\rm freeze}}$ can
be applied to find that $T_{\rm freeze} \simeq 0.5 g_*^{1/6}$ MeV
\cite{kolb}.

Eventually, at a temperature of roughly 0.2 MeV deuterium starts
to form, and very quickly all free neutrons are processed into
$^4$He. The final helium abundance is therefore roughly given by
\begin{equation}
Y_P \simeq \left. \frac{2 n_n/n_p}{1+n_n/n_p} \right|_{T\simeq 0.2
\,\, {\rm MeV}}.
\end{equation}

$n_n/n_p$ is determined by its value at freeze out, roughly by the
condition that $n_n/n_p|_{T=T_{\rm freeze}} \sim
e^{-(m_n-m_p)/T_{\rm freeze}}$.

Since the freeze-out temperature is determined by $g_*$ this in
turn means that $g_*$ can be inferred from a measurement of the
helium abundance. However, since $Y_P$ is a function of both
$\Omega_b h^2$ and $g_*$ it is necessary to use other measurements
to constrain $\Omega_b h^2$ in order to find a bound on $g_*$. One
customary method for doing this has been to use measurements of
primordial deuterium to infer $\Omega_b h^2$ and from that
calculate a bound on $g_*$. Usually such bounds are expressed in
terms of the equivalent number of neutrino species, $N_\nu \equiv
\rho/\rho_{\nu_0}$, instead of $g_*$. The exact value of the bound
is quite uncertain because there are different and mutually
inconsistent measurements of the primordial helium abundance (see
for instance Ref.~\cite{Barger:2003zg,Steigman:2005uz} for a
discussion of this issue). Some of the most recent analyses are
\cite{Barger:2003zg} where a value of $1.7 \leq N_\nu \leq 3.0$
(95\% C.L.) was found, \cite{Serpico:2004gx} which found $-1.14
\leq \Delta N_\nu \leq 0.73$, and \cite{Cyburt:2004yc} which found
that $N_\nu = 3.14^{+0.7}_{-0.65}$ at 68\% C.L. The difference in
these results can be attributed to different assumptions about
uncertainties in the primordial helium abundance. It should be
noted that all these results are consistent with $N_\nu = 3.04$,
the standard model result. If a low helium abundance is assumed
than the BBN bound severely restricts additional relativistic
energy density beyond the standard model prediction. A full extra
neutrino degree of freedom is excluded at high significance.
However, that is not the case when a higher helium abundance is
assumed, and the fact that the bound is dependent on assumptions
should indicate that it is not a level of precision where it can
be completely trusted.

Another interesting parameter which can be constrained by the same
argument is the neutrino chemical potential, $\xi_\nu=\mu_\nu/T$
\cite{Kang:xa,Kohri:1996ke,Orito:2002hf,Ichikawa:2002vn}. At first
sight this looks like it is completely equivalent to constraining
$N_\nu$. However, this is not true because a chemical potential
for electron neutrinos directly influences the $n-p$ conversion
rate. Furthermore, it is crucial to take neutrino flavour
oscillations into account when calculating bounds on the neutrino
chemical potential. This point is discussed in more detail below.

%%%%%%%%%%%%%%%%%%%%%%%%%%%%%%%%%%%%%%%%%%%%%%%%%%%%%%%%%%%%%%%%%%%
\subsection{Can neutrinos recouple?}
\label{sec:rec}
%%%%%%%%%%%%%%%%%%%%%%%%%%%%%%%%%%%%%%%%%%%%%%%%%%%%%%%%%%%%%%%%%%%

Standard model neutrinos interact via exchange of $W$ and $Z$
bosons where the coupling strength at low energy $(E \ll m_{W,Z})$
is given by Fermi's constant
\begin{equation}
G_F = \frac{\pi \alpha}{\sqrt{2}} \frac{1}{\sin^2 \theta_W m_W^2}.
\end{equation}
The $1/m_W^2$ comes from the vector boson propagator. However, at
energies beyond $m_W$ the propagator will have $m_W$ replaced by
momentum exchange $k$. Since this should be roughly equal to the
thermal energy of an average particle in the medium, the effective
interaction rate should go from
\begin{equation}
\Gamma \simeq G_F^2 T^5 \,\,\, (T \ll m_W)
\end{equation}
to
\begin{equation}
\Gamma \simeq \alpha T \,\,\, (T \gg m_W).
\end{equation}
Comparing this with the Hubble parameter which in the radiation
dominated regime is given by Eq.~(\ref{eq:hub}), one finds that
\begin{equation}
\frac{\Gamma}{H} \propto T^{-1}.
\end{equation}
At high temperatures $\Gamma/H$ will be less than 1 and neutrinos
will be out of equilibrium. The temperatures where this happens is
roughly $T/m_{\rm Pl} \simeq \alpha/g_*^{1/2} \sim 10^{-3}$ which
is far beyond the range where the standard model applies.

However, even if the above example is therefore not physically
relevant it shows that neutrinos which interact via exchange of
massless particles will always have an interaction rate (for $2
\leftrightarrow 2$ processes) which is
\begin{equation}
\Gamma = \beta T,
\end{equation}
where $\beta$ is a dimensionless quantity. Therefore such
interactions will always drive neutrinos {\it toward} thermal
equilibrium at late times instead of out of equilibrium.

\begin{figure}[htbp]
\begin{center}
\epsfig{file=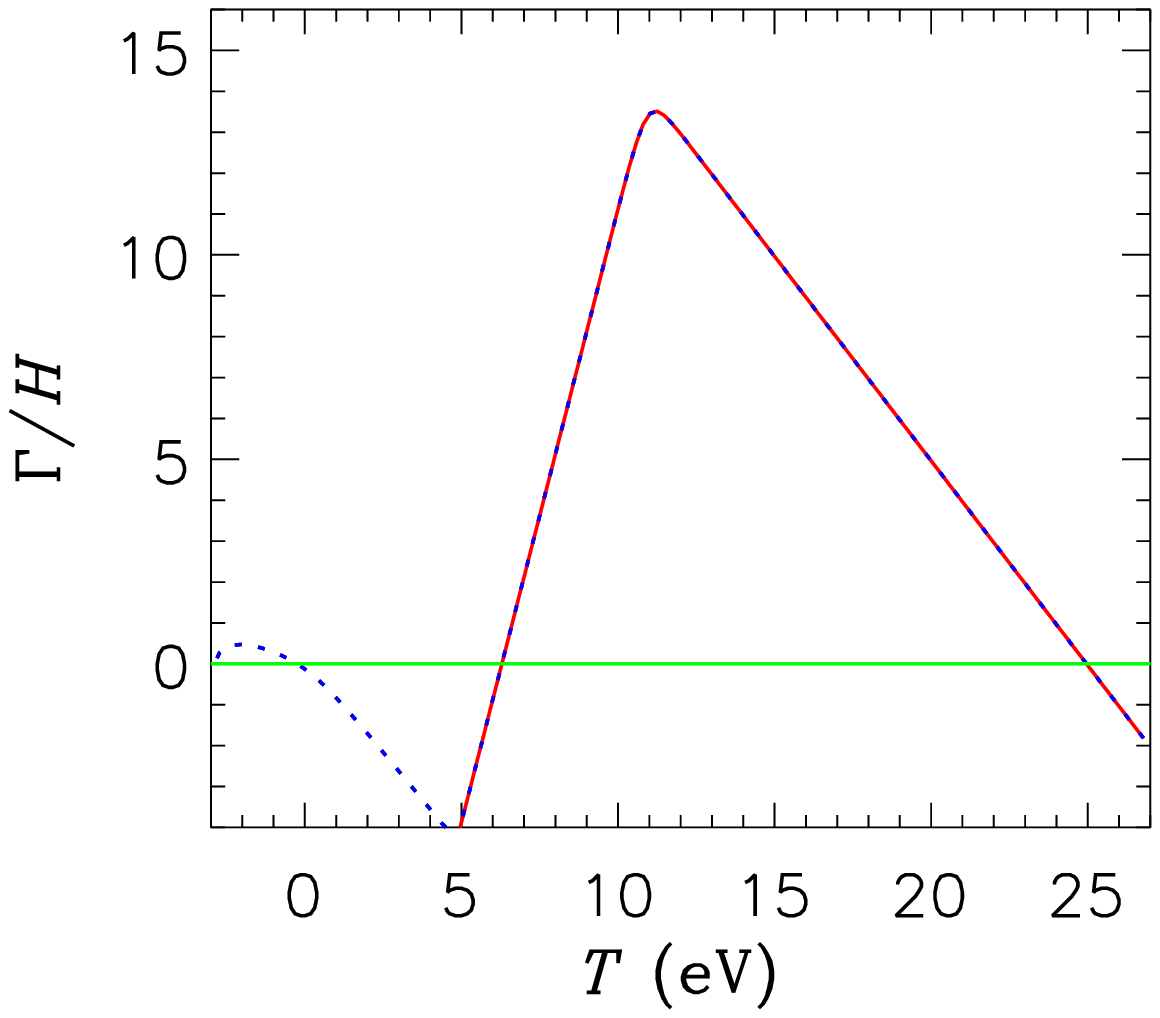,width=0.8\textwidth}
\end{center}
\bigskip
\caption{\label{fig:nueq} The interaction rate for standard a
standard model neutrino is shown in red. The blue curve is for a
0.003 eV neutrino coupled to a massless scalar with dimensionless
coupling $g=3 \times 10^{-7}$.}
\end{figure}

One example which has received significant attention recently is
the possibility that neutrinos couple to a massless scalar or
pseudoscalar particle. If the interaction is sufficiently strong
the neutrinos will come into equilibrium after BBN but before the
present and consequently they will annihilate and disappear when
$T \sim m_\nu$
\cite{Beacom:2004yd,Hannestad:2004qu,Sawyer:2006ju}. In
Fig.~\ref{fig:nueq} we show $\Gamma/H$ for massless standard model
neutrinos. As can be seen, neutrinos decouple from thermal
equilibrium when $T \sim 1$ MeV. However for $T \gg m_Z$ the cross
section drops and neutrinos are out of equilibrium at very early
times. In the same figure we also show what happens if neutrinos
couple to a light scalar, in which case they come into equilibrium
at late times.

\subsection{The effect of oscillations}

In the previous section the one-particle distribution function,
$f$, was used to describe neutrino evolution. However, for
neutrinos the mass eigenstates are not equivalent to the flavour
eigenstates because neutrinos are mixed. Therefore the evolution
of the neutrino ensemble is not in general described by the three
scalar functions, $f_i$, but rather by the evolution of the
neutrino density matrix, $\rho \equiv \psi \psi^\dagger$, the
diagonal elements of which correspond to $f_i$.

For three-neutrino oscillations the formalism is quite
complicated. However, the difference in $\Delta m_{12}$ and
$\Delta m_{23}$, as well as the fact that $\sin 2 \theta_{13} \ll
1$ means that the problem effectively reduces to a $2 \times 2$
oscillation problem in the standard model. The effect of
oscillations on the decoupling of neutrinos was discussed in
\cite{osc,Mangano:2005cc}. A detailed account of the physics of
neutrino oscillations in the early universe is outside the scope
of the present review, however an excellent and very thorough
discussion of this topic can be found in Ref.~\cite{dolgov}.

Without oscillations it is possible to compensate a very large
chemical potential for muon and/or tau neutrinos with a small,
negative electron neutrino chemical potential \cite{Kang:xa}.
However, since neutrinos are almost maximally mixed a chemical
potential in one flavour can be shared with other flavours, and
the end result is that during BBN all three flavours have almost
equal chemical potential
\cite{lunardini,Pastor:2001iu,Dolgov:2002ab,Abazajian:2002qx,Wong:2002fa}.
This in turn means that the bound on $\nu_e$ applies to all
species. The most recent bound on the neutrino asymmetry from BBN
comes from \cite{Serpico:2005bc}

\begin{equation}
-0.04 \leq \xi_i = \frac{|\eta_i|}{T} \leq 0.07
\end{equation}
for $i=e,\mu,\tau$.

The bound assumes complete flavour equilibration during BBN, which
with the measured mixing angles and mass differences is a fairly
good approximation. In models where sterile neutrinos are present
even more remarkable oscillation phenomena can occur. However, I
do not discuss this possibility further, except for the
possibility of sterile neutrino warm dark matter, and instead
refer to the review \cite{dolgov}.

\subsection{Low reheating temperature and neutrinos}

In most models of inflation the universe enters the normal,
radiation dominated epoch at a reheating temperature, $T_{\rm
RH}$, which is of order the electroweak scale or higher. However,
in principle it is possible that this reheating temperature is
much lower, of order MeV. This possibility has been studied many
times in the literature, and a very general bound of $T_{\rm RH}
\gtrsim 1$ MeV has been found
\cite{Kawasaki:1999na,Kawasaki:2000en,Giudice:2000ex,Giudice:2000dp}

This very conservative bound comes from the fact that the light
element abundances produced by big bang nucleosynthesis disagree
with observations if the universe if matter dominated during BBN.
However, a somewhat more stringent bound can be obtained by
looking at neutrino thermalization during reheating. If a scalar
particle is responsible for reheating then direct decay to
neutrinos is suppressed because of the necessary helicity flip.
This means that if the reheating temperature is too low neutrinos
never thermalize. If this is the case then BBN predicts the wrong
light element abundances. However, even if the heavy particle has
a significant branching ratio into neutrinos there are problems
with BBN. The reason is that neutrinos produced in decays are born
with energies which are much higher than thermal. If the reheating
temperature is too low then a population of high energy neutrinos
will remain and also lead to conflict with observed light element
abundances. Recent analyses have shown that in general the
reheating temperature cannot be below a few MeV
\cite{Hannestad:2004px} (see also \cite{Ichikawa:2005vw}).

%%%%%%%%%%%%%%%%%%%%%%%%%%%%%%%%%%%%%%%%%%%%%%%%%%%%%%%%%%%%%%%%%%%%%%
\section{Cosmological data} %%%%%%%%%%%%%%%%%%%%%%%%%%%%%%%%%%%%%%%%%%
%%%%%%%%%%%%%%%%%%%%%%%%%%%%%%%%%%%%%%%%%%%%%%%%%%%%%%%%%%%%%%%%%%%%%%

Even though BBN considerations do provide interesting bounds on
neutrino physics attention has shifted towards observations of the
late-time Universe. In recent years observations of physics after
recombination has provided very strong constraint on many
cosmological parameters and also on neutrino physics. In this
section I review the current status of cosmological observations.

\paragraph{Large Scale Structure (LSS) --}

The Sloan Digital Sky Survey (SDSS) has measured redshifts of
close to 1 million galaxies, providing the so far most detailed
map of the large scale structure of the Universe
\cite{Tegmark:2003uf,Tegmark:2003ud}. The SDSS data has been used
extensively for cosmological parameter estimation, including
neutrino mass bounds. The somewhat smaller 2dFGRS (2~degree Field
Galaxy Redshift Survey) \cite{2dFGRS} has also been used for the
same purpose. Both surveys provide a precise measurement of the
galaxy-galaxy power spectrum, $P_g(k)$, down to $k \sim 0.01
h$/Mpc. In order to derive the matter power spectrum, $P_m(k)$,
from this it is necessary to know the bias parameter, $b$, defined
as $b^2(k) = P_g(k)/P_m(k)$. In general the bias parameter is
scale dependent and a function of the complicated hydrodynamics of
non-linear structure formation. However, if only data on scales
larger than about $k \sim 0.15 h$/Mpc is used the bias parameter
can be taken to be a constant. In principle the bias parameter can
be determined from higher-order correlations (see for instance
\cite{Verde:2001sf,Seljak:2004sj}), and it does provide a
stringent constraint on the neutrino mass \cite{Seljak:2004xh}.
However the systematic error on the bias parameter does rely on
the assumption that the non-linear aspects of structure formation
are well understood, an assumption which is not necessarily
justified at present. Therefore the bound on neutrino mass coming
from the use of bias constraints should probably be regarded as
somewhat less robust than bounds relying only on the power
spectrum shape.

\paragraph{Cosmic Microwave Background (CMB) --}

Fluctuations in the CMB were first measured by the COBE satellite
in 1992 \cite{Smoot:1992td}. The temperature fluctuations in the
CMB can be conveniently decomposed into spherical harmonics
\begin{equation}
\frac{\Delta T}{T} (\theta,\phi) = \sum_{lm}
a_{lm}Y_{lm}(\theta,\phi).
\end{equation}
From the $a_{lm}$ coefficients one can construct the angular power
spectrum
\begin{equation}
C_{T,l} \equiv \langle |a_{lm}|^2 \rangle,
\end{equation}
where $\langle ... \rangle$ denotes an ensemble average. Since
only one realization of the underlying distribution is available
it is in practise replaced by an angular average corresponding to
averaging over $2l+1$ $m$-modes for each $l$.

The dominant scattering mechanism for photons around the apoch of
recombination is Thomson scattering on free electrons. Since
Thomson scattering polarizes light any anisotropy in the electron
distribution leads to a net polarization anisotropy in the CMB.
Like the temperature anisotropy, polarization can be written in
terms of angular power spectra. Because there is no circular
polarization component in the CMB there are only two independent
components in the polarization tensor. The usual decomposition is
in terms of curl-free $(E)$ and a curl $(B)$ component which
yields four independent power spectra: $C_{T,l}$, $C_{E,l}$,
$C_{B,l}$, and the $T$-$E$ cross-correlation $C_{TE,l}$. There is
no cross-correlation between $B$ and $E,T$ because of different
parity.

The WMAP experiment has reported data on $C_{T,l}$ and $C_{TE,l}$
as described in
Refs.~\cite{Spergel:2003cb,Bennett:2003bz,Kogut:2003et,%
Hinshaw:2003ex,Verde:2003ey}. Foreground contamination has already
been subtracted from their published data. However, it is well
known that the WMAP 1-year data does exhibit some notable
anomalies. First, there are several ''glitches'' in the power
spectrum, which are presumably due to incomplete foreground
removal. Second, there is a marked suppression and alignment of
the low-$w$ multipoles. There has been substantial discussion of
the nature of the low-$l$ anomaly, but with no clear conclusion.
Very importantly neither of the two anomalies have any influence
at all on the study of neutrino properties. The reason is that
neutrino physics affects only scales smaller than roughly the
horizon size at recombination, and that changes to neutrino
physics cannot produce sharp features in angular CMB power
spectrum. Therefore, any conclusion on neutrinos is likely to
hold, independent of the possible nature of the anomalies.

In addition to the WMAP data there are many independent
measurements of the spectrum on smaller scales by ground based or
balloon borne experiments. The most important of these at present
is the Boomerang experiment
\cite{Jones:2005yb,Piacentini:2005yq,Montroy:2005yx} which has
measured significantly smaller scales than WMAP. Furthermore, this
experiment has been the first to measure $C_{E,l}$.

\paragraph{The baryon acoustic peak (BAO) --}

In the early Universe, prior to recombination, baryons and photons
undergo acoustic oscillations. These oscillations are not only
detectable in the CMB, but in principle also in the LSS power
spectrum. However, since baryons are only a subdominant component
early structure formation is dominated by cold dark matter, and
the oscillations are merely a small amplitude modulation of the
power spectrum. In terms of the real-space correlation function
this modulation corresponds to a well-defined peak, located at a
scale of roughly 100$h^{-1}$ Mpc. This peak was first measured
using the bright red galaxies of the SDSS
\cite{Eisenstein:2005su}. The position of the peak provides an
accurate measurement of the angular diameter distance out to
redshift of roughly 0.35. This in turn provides a constraint on
$\Omega_m$, $\Omega_{\rm DE}$, and $w$. Furthermore the constraint
shows relatively little sensitivity to other parameters and seems
less prone to systematics than measurements of the power spectrum
amplitude.

\paragraph{The Lyman-$\alpha$ forest --}

Measurements of the flux power spectrum of the Lyman-$\alpha$
forest has been used to measure the matter power spectrum on small
scales at large redshift. This has been done using high resolution
Keck or VLT spectra \cite{Croft:2000hs,Kim:2003qt}. By far the
largest sample of spectra comes from the SDSS survey, even though
the spectral resolution is substantially poorer than in for Keck
or VLT. In Ref.~\cite{mcdonald} this data was carefully analyzed
and used to constrain the linear matter power spectrum. The
derived amplitude is $\Delta^2 (k=0.009 \, {\rm km \, s}^{-1},
z=3) = 0.452^{+0.07}_{-0.06}$ and the effective spectral index is
$n_{\rm eff} = -2.321^{+0.06}_{-0.05}$. Even though the higher
resolution data extend almost an order of magnitude higher in $k$
the smallest scales are likely to be completely dominated by
systematics related to non-linearities so that relatively little
additional information would be gained. In order to probe the mass
of light neutrinos the SDSS data is therefore optimal. On the
other hand the small scale data is important for testing models
for warm dark matter, as will be discussed later.

The quote SDSS result has been derived using a very elaborate
model for the local intergalactic medium, including full N-body
simulations. It has been shown that using the Lyman-$\alpha$ data
does strengthen the bound on neutrino mass significantly. However
the question remains as to the level of systematic uncertainty in
the result. Especially the amplitude of the matter power spectrum
is quite sensitive to model assumptions. Therefore the same
caution should be applied to bounds using this result as to the
bound from LSS data using bias constraints, as discussed above.

\paragraph{Type Ia supernovae --}

Measurements of distant type Ia supernovae provide a precise
measurement of the luminosity distance $d_L$ as a function of
redshift \cite{Riess:1998cb,Perlmutter:1998np}. This in turn
provides an important constraint on $H(z)$ and therefore on
$\Omega_m$, $\Omega_{\rm DE}$, and $w_{\rm DE}$. The most recent
and precise data set comes from the SuperNova Legacy Survey (SNLS)
which has published data for 71 supernovae \cite{snls}. Another
data set which has been extensively used is the Riess et al.
''gold'' data set \cite{Riess:2004} which consists of 157
supernovae, measured using both HST and ground based telescopes.

\paragraph{Other data --}

Apart from the above mentioned data the measurement of the Hubble
constant by the HST Hubble Key Project, $H_0=72 \pm 8~{\rm
km}~{\rm s}^{-1}~{\rm Mpc}^{-1}$ \cite{Freedman:2000cf} is
important for cosmological parameter analyses.

%%%%%%%%%%%%%%%%%%%%%%%%%%%%%%%%%%%%%%%%%%%%%%%%%%%%%%%%%%%%%%%%%%%%%%
\section{Neutrino Dark Matter} %%%%%%%%%%%%%%%%%%%%%%%%%%%%%%%%%%%%%%%
%%%%%%%%%%%%%%%%%%%%%%%%%%%%%%%%%%%%%%%%%%%%%%%%%%%%%%%%%%%%%%%%%%%%%%

Neutrinos are a source of dark matter in the present day universe
simply because they contribute to $\Omega_m$. The present
temperature of massless standard model neutrinos is $T_{\nu,0} =
1.95 \, K = 1.7 \times 10^{-4}$ eV, and any neutrino with $m \gg
T_{\nu,0}$ behaves like a standard non-relativistic dark matter
particle.

The present contribution to the matter density of $N_\nu$ neutrino
species with standard weak interactions is given by
\begin{equation}
\Omega_\nu h^2 = N_\nu \frac{m_\nu}{93.8 \, {\rm eV}}
\end{equation}
Just from demanding that $\Omega_\nu \leq 1$ one finds the bound
\cite{Gershtein:gg,Cowsik:gh}
\begin{equation}
m_\nu \lesssim \frac{46 \, {\rm eV}}{N_\nu} \label{eq:mnu}
\end{equation}

\subsection{The Tremaine-Gunn bound}

If neutrinos are the main source of dark matter, then they must
also make up most of the galactic dark matter. However, neutrinos
can only cluster in galaxies via energy loss due to gravitational
relaxation since they do not suffer inelastic collisions. In
distribution function language this corresponds to phase mixing of
the distribution function \cite{Tremaine:we}. By using the theorem
that the phase-mixed or coarse grained distribution function must
explicitly take values smaller than the maximum of the original
distribution function one arrives at the condition
\begin{equation}
f_{\rm CG} \leq f_{\nu,{\rm max}} = \frac{1}{2}
\end{equation}
Because of this upper bound it is impossible to squeeze neutrino
dark matter beyond a certain limit \cite{Tremaine:we}. For the
Milky Way this means that the neutrino mass must be larger than
roughly 25 eV {\it if} neutrinos make up the dark matter. For
irregular dwarf galaxies this limit increases to 100-300 eV
\cite{Madsen:mz,salucci}, and means that standard model neutrinos
cannot make up a dominant fraction of the dark matter. This bound
is generally known as the Tremaine-Gunn bound.

Note that this phase space argument is a purely classical
argument, it is not related to the Pauli blocking principle for
fermions (although, by using the Pauli principle $f_\nu \leq 1$
one would arrive at a similar, but slightly weaker limit for
neutrinos). In fact the Tremaine-Gunn bound works even for bosons
if applied in a statistical sense \cite{Madsen:mz}, because even
though there is no upper bound on the fine grained distribution
function, only a very small number of particles reside at low
momenta (unless there is a condensate). Therefore, although the
exact value of the limit is model dependent, limit applies to any
species that was once in thermal equilibrium. A notable
counterexample is non-thermal axion dark matter which is produced
directly into a condensate.

A very interesting direct example of the Tremaine-Gunn bound was
studied in \cite{Ringwald:2004np}. Here, neutrino clustering in
evolving cold dark matter halos was studied using the Boltzmann
equation. The problem was made tractable by assuming the
backreaction, i.e.\ that the CDM halos are not affected by
neutrinos. The results clearly show that there is a maximum in the
coarse grained distribution function of $1/2$, as expected. The
calculation was extended to bosons in \cite{Hannestad:2005bt},
where no such upper bound was found. In fact bosons can have an
average density several times higher than fermions with equal mass
inside dark matter halos, an effect which in principle might be
used to distinguish fermionic and bosonic hot dark matter.

\subsection{Neutrino hot dark matter}

A much stronger upper bound on the neutrino mass than the one in
Eq.~(\ref{eq:mnu}) can be derived by noticing that the thermal
history of neutrinos is very different from that of a WIMP because
the neutrino only becomes non-relativistic very late.

In an inhomogeneous universe the Boltzmann equation for a
collisionless species is \cite{MB}
\begin{equation}
L[f] = \frac{Df}{D\tau} = \frac{\partial f}{\partial \tau}
 + \frac{dx^i}{d\tau}\frac{\partial f}{\partial x^i} +
\frac{dq^i}{d\tau}\frac{\partial f}{\partial q^i} = 0,
\end{equation}
where $\tau$ is conformal time, $d \tau = dt/a$, and $q^i = a p^i$
is comoving momentum. The second term on the right-hand side has
to do with the velocity of the distribution in a given spatial
point and the third term is the cosmological momentum redshift.

Following Ma and Bertschinger \cite{MB} this can be rewritten as
an equation for $\Psi$, the perturbed part of $f$
\begin{equation}
f(x^i,q^i,\tau) = f_0(q) \left[ 1 + \Psi(x^i,q^i,\tau) \right]
\end{equation}

In synchronous gauge that equation is

\begin{equation}
\frac{1}{f_0}[f] = \frac{\partial \Psi}{\partial \tau} + i
\frac{q}{\epsilon} \mu \Psi + \frac{d \ln f_0}{d \ln q}
\left[\dot{\eta}-\frac{\dot{h}+6\dot{\eta}} {2} \mu^2 \right] =
\frac{1}{f_0} C[f],
\end{equation}
where $q^j = q n^j$, $\mu \equiv n^j \hat{k}_j$, and $\epsilon =
(q^2 + a^2 m^2)^{1/2}$. $k^j$ is the comoving wavevector. $h$ and
$\eta$ are the metric perturbations, defined from the perturbed
space-time metric in synchronous gauge \cite{MB}
\begin{equation}
ds^2 = a^2(\tau) [-d\tau^2 + (\delta_{ij} + h_{ij})dx^i dx^j],
\end{equation}
\begin{equation}
h_{ij} = \int d^3 k e^{i \vec{k}\cdot\vec{x}}\left(\hat{k}_i
\hat{k}_j h(\vec{k},\tau) +(\hat{k}_i \hat{k}_j - \frac{1}{3}
\delta_{ij}) 6 \eta (\vec{k},\tau) \right).
\end{equation}

Expanding this in Legendre polynomials one arrives at a set of
hierarchy equations
\begin{eqnarray}
\dot{\delta} & = & -\frac{4}{3} \theta - \frac{2}{3} \dot h \nonumber \\
\dot{\theta} & = & k^2\left(\frac{\delta}{4} - \sigma \right) \nonumber \\
2 \dot{\sigma} & = & \frac{8}{15} \theta - \frac{3}{15} k F_3
+ \frac{4}{15} \dot h + \frac{8}{5} \dot{\eta} \nonumber \\
\dot{F}_l & = & \frac{k}{2l+1} \left(l F_{l-1} - (l+1) F_{l+1}
\right)
\end{eqnarray}
For subhorizon scales ($\dot h = \dot \eta = 0$) this reduces to
the form
\begin{eqnarray}
\dot{\delta} & = & -\frac{4}{3} \theta \nonumber \\
\dot{\theta} & = & k^2\left(\frac{\delta}{4} - \sigma \right) \nonumber \\
2 \dot{\sigma} & = & \frac{8}{15} \theta - \frac{3}{15} k F_3 \nonumber \\
\dot{F}_l & = & \frac{k}{2l+1} \left(l F_{l-1} - (l+1) F_{l+1}
\right)
\end{eqnarray}

One should notice the similarity between this set of equations and
the evolution hierarchy for spherical Bessel functions. Indeed the
exact solution to the hierarchy is
\begin{equation}
F_{l}(k \tau) \sim j_l(k \tau)
\end{equation}
This shows that the solution for $\delta$ is an exponentially
damped oscillation. On small scales, $k > \tau$, perturbations are
erased.

This in intuitively understandable in terms of free-streaming.
Indeed the Bessel function solution comes from the fact that
neutrinos are considered massless. In the limit of CDM the
evolution hierarchy is truncated by the fact that $\theta=0$, so
that the CDM perturbation equation is simply $\dot \delta = -\dot
h/2$. For massless particles the free-streaming length is $\lambda
= c \tau$ which is reflected in the solution to the Boltzmann
hierarchy. Of course the solution only applies when neutrinos are
strictly massless. Once $T \sim m$ there is a smooth transition to
the CDM solution. Therefore the final solution can be separated
into two parts: 1) $k > \tau(T=m)$: Neutrino perturbations are
exponentially damped 2) $k < \tau(T=m)$: Neutrino perturbations
follow the CDM perturbations. Calculating the free streaming
wavenumber in a flat CDM cosmology leads to the simple numerical
relation (applicable only for $T_{\rm eq} \gg m \gg T_0$)
\cite{kolb}

\begin{equation}
\lambda_{\rm FS} \sim \frac{20~{\rm Mpc}}{\Omega_x h^2}
\left(\frac{T_x}{T_\nu}\right)^4 \left[1+\log \left(3.9
\frac{\Omega_x h^2}{\Omega_m h^2} \left(\frac{T_\nu}{T_x}\right)^2
\right)\right]\,. \label{eq:freestream}
\end{equation}

\begin{figure}[htbp]
\begin{center}
\epsfig{file=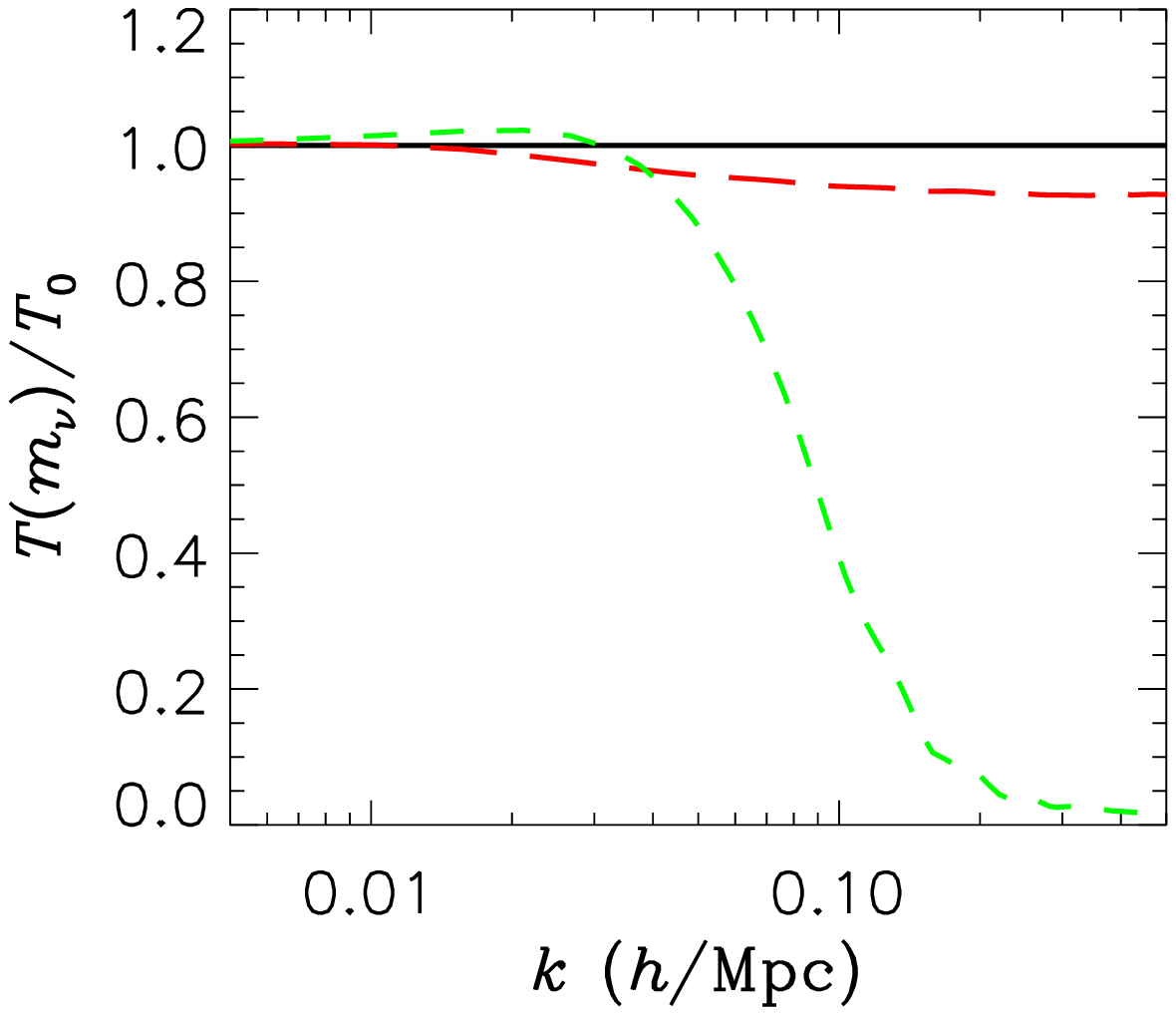,width=0.8\textwidth}
\end{center}
\bigskip
\caption{\label{fig:nutrans} The transfer function $T(k,t=t_0)$
for various different neutrino masses. The solid (black) line is
for $m_\nu=0$, the long-dashed for $m_\nu = 0.3$ eV, and the
dashed for $m_\nu=1$ eV.}
\end{figure}

In Fig.~\ref{fig:nutrans} I have plotted transfer functions for
various different neutrino masses in a flat $\Lambda$CDM universe
$(\Omega_m+\Omega_\nu+\Omega_\Lambda=1)$. The parameters used were
$\Omega_b = 0.04$, $\Omega_{\rm CDM} = 0.26 - \Omega_\nu$,
$\Omega_\Lambda = 0.7$, $h = 0.7$, and $n=1$.

When measuring fluctuations it is customary to use the power
spectrum, $P(k,\tau)$, defined as
\begin{equation}
P(k,\tau) = |\delta|^2(\tau).
\end{equation}
The power spectrum can be decomposed into a primordial part,
$P_0(k)$, and a transfer function $T(k,\tau)$,
\begin{equation}
P(k,\tau) = P_0(k) T(k,\tau).
\end{equation}
The transfer function at a particular time is found by solving the
Boltzmann equation for $\delta(\tau)$.

At scales much smaller than the free-streaming scale the present
matter power spectrum is suppressed roughly by the factor
\cite{Hu:1997mj}
\begin{equation}
\frac{\Delta P(k)}{P(k)} = \frac{\Delta
T(k,\tau=\tau_0)}{T(k,\tau=\tau_0)}\simeq -8
\frac{\Omega_\nu}{\Omega_m},
\end{equation}
as long as $\Omega_\nu \ll \Omega_m$. The numerical factor 8 is
derived from a numerical solution of the Boltzmann equation, but
the general structure of the equation is simple to understand. At
scales smaller than the free-streaming scale the neutrino
perturbations are washed out completely, leaving only
perturbations in the non-relativistic matter (CDM and baryons).
Therefore the {\it relative} suppression of power is proportional
to the ratio of neutrino energy density to the overall matter
density. Clearly the above relation only applies when $\Omega_\nu
\ll \Omega_m$, when $\Omega_\nu$ becomes dominant the spectrum
suppression becomes exponential as in the pure hot dark matter
model. This effect is shown for different neutrino masses in
Fig.~\ref{fig:nutrans}.

The effect of massive neutrinos on structure formation only
applies to the scales below the free-streaming length. For
neutrinos with masses of several eV the free-streaming scale is
smaller than the scales which can be probed using present CMB data
and therefore the power spectrum suppression can be seen only in
large scale structure data. On the other hand, neutrinos of sub-eV
mass behave almost like a relativistic neutrino species for CMB
considerations. The main effect of a small neutrino mass on the
CMB is that it leads to an enhanced early ISW effect. The reason
is that the ratio of radiation to matter at recombination becomes
larger because a sub-eV neutrino is still relativistic or
semi-relativistic at recombination.

%%%%%%%%%%%%%%%%%%%%%%%%%%%%%%%%%%%%%%%%%%%%%%%%%%%%%%%%%%%%%%%
\subsection{Neutrino mass bounds}
%%%%%%%%%%%%%%%%%%%%%%%%%%%%%%%%%%%%%%%%%%%%%%%%%%%%%%%%%%%%%%%

\subsubsection{Parameter estimation methodology}

Because massive neutrinos affect structure formation it is
possible to constrain their mass using a combination of
cosmological data. The standard approach to cosmological parameter
estimation is to use Bayesian statistics which provides a very
simple method for incorporating prior information on parameters
from other sources. Using the prior probability distribution it is
then possible to calculate the posterior distribution and from
that to derive confidence limits on parameters. There are standard
packages such as CosmoMC \cite{cosmomc}, a likelihood calculator
based on the Markov Chain Monte Carlo method
\cite{Christensen:2001gj,Lewis:2002ah}, available for this
purpose.

It should be noted here that confidence limits derived using
Bayesian statistics in this way are often much more stringent than
implied by frequentist statistics, in which one calculates the
confidence with which a given hypothesis can be excluded. The main
reason for this is that observables in our Universe is a single
realization of an underlying distribution. There is no possibility
for producing new data to test the hypothesis and therefore the
frequentist approach in many cases does not fully use the
information provided by the given data. Much more detailed
discussions of this point can be found in \cite{liddle,trotta}. In
the remainder of this section we discuss only results derived
using the Bayesian method.

Independent of the statistical method used results will in general
depend on the number of parameters used to fit the data. Because
of parameter degeneracies bounds on a given parameter will in
general get weaker if more parameters are used in the fit.
However, the question remains as to how many parameters should
plausibly be included. If nothing is known a priori about the
cosmological model the best-fit $\chi^2$ for $N$ parameters should
in general be substantially lower for $N+1$ parameters if the
additional parameter is to be included. This point has been
extensively discussed in \cite{liddle}, and if used at face value
indicates that the vanilla $\Lambda$CDM model with $n=1$ is the
preferred model. This model assumes spatial flatness and uses only
the following free parameters: $\Omega_m$, the matter density,
$\Omega_b$, the baryon density, $H_0$, the Hubble parameters,
$\tau$, the optical depth to reionization, and $A$, the amplitude
of primordial density fluctuations. It furthermore assumes that
the primordial perturbations are purely adiabatic in nature.

However, even though the best fit $\chi^2$ for neutrino mass is
for $m_\nu = 0$ that certainly does not mean the neutrino masses
should not be included in parameter fitting. The neutrino mass is
known from oscillation experiments to be non-zero and therefore
must be included in parameter estimation.

Furthermore almost all studies of constraints on non-standard
cosmological parameters assume the minimal $\Lambda$CDM model as
the benchmark and then include the non-standard parameters
specific to the given study. This approach is dangerous because
there could easily be several non-standard parameters which
produce a much better fit when combined. It is important to study
more general models, especially since numerical techniques have
now made it feasible to use many more free parameters. Below we
will discuss one case where including two non-standard parameters
can lead to completely changed conclusions about neutrino mass
bounds.

\subsubsection{Current bounds}\footnote{Another recent review
discussing neutrino mass bounds is \protect\cite{Elgaroy:2004rc}}

With the WMAP data alone sensitivity to the neutrino mass is
relatively limited. In a standard likelihood analysis for the 6
parameters of the minimal $\Lambda$CDM model plus neutrino mass we
find an upper bound of $\sum m_\nu < 2.1$ eV (95\% C.L.). This
value is consistent with that derived in other recent analyses
\cite{Ichikawa:2004zi,julien}.

Including LSS data the bound can be improved. However, if there is
no information on the normalization of the large scale structure
power spectrum (the bias parameter), the improvement is relatively
modest. For the same model as above the bound for WMAP+SDSS is 1.8
eV \cite{Tegmark:2003ud}.

Once other data, such as type Ia supernova measurements, are
included the bound can be improved significantly. A very stringent
constraint comes from including information on the normalization
of the matter power spectrum from measurements of the
Lyman-$\alpha$ forest. In \cite{Seljak:2004xh} a bound of $\sum
m_\nu < 0.42$ eV (95\% C.L.) was derived using WMAP, SDSS, SNI-a,
and Lyman-$\alpha$ data for the standard $\Lambda$CDM model with
neutrino mass (\cite{Fogli:2004as} found an almost identical
result). However, the uncertainty related to the modelling of the
intergalactic medium at high redshift does make the bound model
dependent.

Another very interesting bound can be derived when the measurement
of the baryon acoustic oscillation (BAO) peak
\cite{Eisenstein:2005su} is included \cite{newp}. Even for a very
general model with 11 free parameters (adding the number of
neutrino species $N_\nu$, a possible running of the spectral
index, $\alpha_s$, and the dark energy equation of state, $w$) the
upper bound was found to be 0.48 eV (95\% C.L.) \cite{newp}. This
is quite interesting because without the BAO data it was shown in
\cite{Hannestad:2005gj} that there is a severe degeneracy between
$\sum m_\nu$ and $w$. For CMB, LSS and SNI-a data the bound on
$\sum m_\nu$ can be relaxed by as much as a factor of 3 when $w$
is allowed to vary. This degeneracy is broken by adding the BAO
data because it provides a tight relation between $\Omega_m$ and
$w$. The result is also encouraging because the systematic
uncertainty in the BAO measurement is likely to be much smaller
than in the Lyman-$\alpha$ forest data. Furthermore the derived
best fit model is almost exactly identical to the best fit model
found in \cite{Seljak:2004xh}, indicating that the Lyman-$\alpha$
and BAO data are completely consistent.

Finally, if both BAO and Lyman-$\alpha$ data is combined the bound
strengthens to $\sum m_\nu < 0.30$ eV (95\% C.L.) \cite{newp}, but
with the same caveat with regards to the Lyman-$\alpha$ data as
before.

In conclusion the current bound on the sum of neutrino masses can
be in the range between 0.3 and 2 eV, depending on the data and
parameters used. In general: \vspace*{0.2cm}\\ a) The bound on
$\sum m_\nu$ for the minimal $\Lambda$CDM plus neutrino mass (8
parameters in total) is in the 0.5-0.6 eV range if CMB, LSS, and
SNI-a data is used.
\vspace*{0.2cm}\\
b) This bound can be relaxed by a large factor when more
parameters, such as $w$ and $N_\nu$ are included, showing that the
bound is not robust.
\vspace*{0.2cm}\\
c) Even including several new parameters the bound can be pushed
below 0.5 eV when additional data is added to the basic CMB, LSS,
and SNI-a data. Either baryon acoustic oscillation or
Lyman-$\alpha$ forest data can be used, with almost identical
result. Especially the BAO data seem to be almost free of
systematics, and consequently the bound seems to be robustly in
the 0.5 eV range.
\vspace*{0.2cm}\\
For those interested an (incomplete) list of other studies of
neutrino mass bounds from cosmological data can be found in
\cite{Elgaroy:2003yh,Hannestad:2003ye,Crotty:2004gm,%
Hannestad:2002xv,Allen:2003pt,Barger:2003vs}.

In Table \ref{table:fits} various upper bounds on neutrino mass
are summarized. The cases 1-4 are taken from \cite{newp} and show
quite clearly that results are strongly dependent on data and
parameters used. Also shown in this table is a selection of other
recent results, all based on the standard $\Lambda$CDM plus
neutrino parameter set.

\begin{figure}[htbp]
\begin{center}
\hspace*{-1cm}\epsfig{file=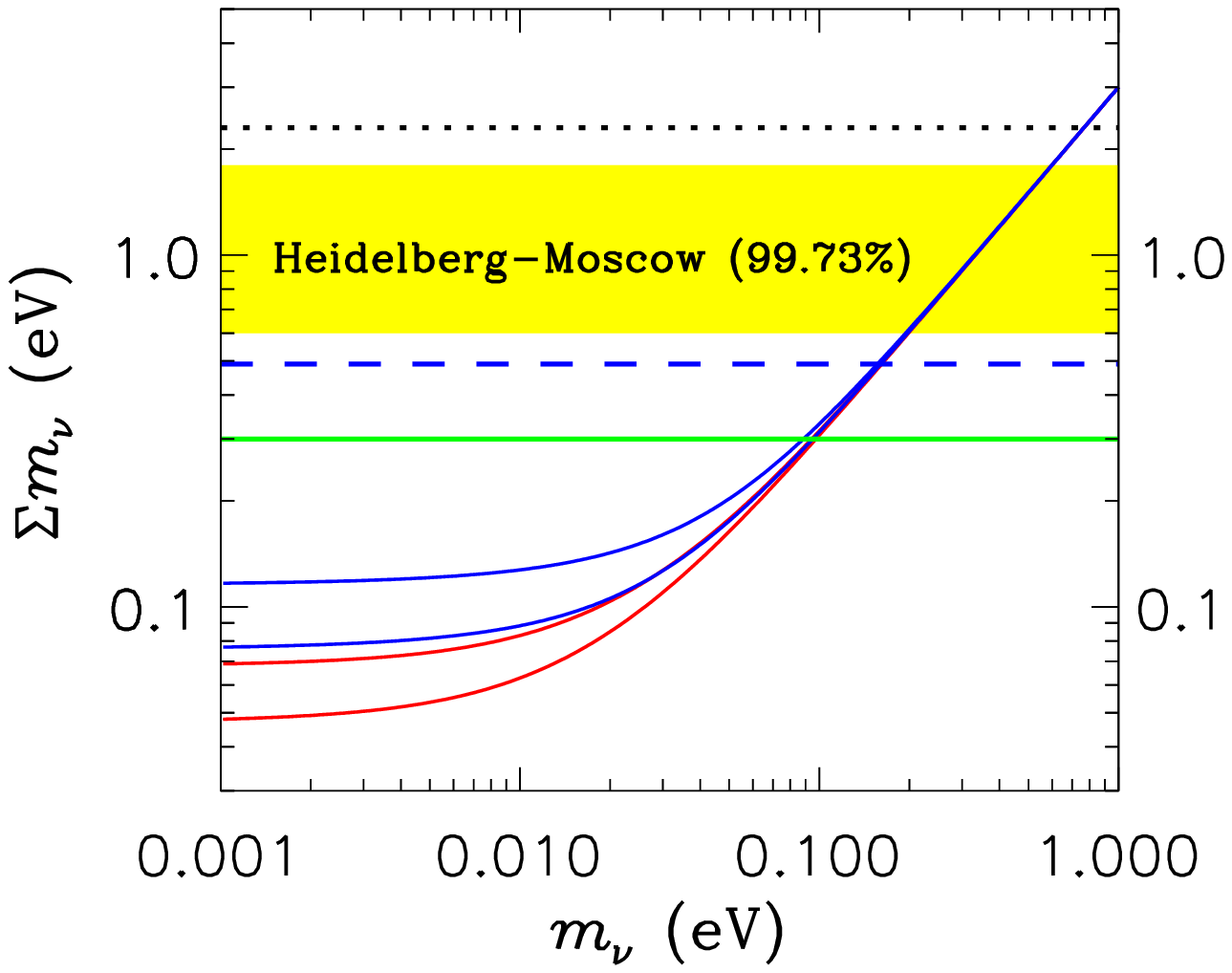,width=0.8\textwidth}
\end{center}
\bigskip
\caption{The 95\% upper bounds on $\sum m_\nu$ for the different
cases presented in Table \protect\ref{table:fits}. The horizontal
axis is the mass of the lightest neutrino mass eigenstate. The set
of full red lines correspond to the normal neutrino mass hierarchy
and the blue to the inverted mass hierarchy. The horizontal yellow
band is the claimed result from the Heidelberg-Moscow experiment.
Note that the exact allowed region from this experiment is
somewhat uncertain due to uncertainties in the nuclear matrix
element. Figure taken from \cite{newp}.} \label{fig:numass}
\end{figure}

In Fig.~\ref{fig:numass} we show the upper bound on the sum of
neutrino masses for various cases which are presented in Table
\ref{table:fits}. Also shown in this figure is the sum of neutrino
masses as a function of the mass of the lightest mass eigenstate.
The red band is for the normal hierarchy and the blue is for the
inverted hierarchy. All bounds were calculated assuming that the
neutrino mass is shared equally between all species. As can be
seen from the figure that assumption is clearly justified at
present. In the same figure we also show the best fit region for
the claimed detection of neutrinoless double beta decay by the
Heidelberg-Moscow experiment
\cite{Klapdor-Kleingrothaus:2001ke,Klapdor-Kleingrothaus:2004wj,%
VolkerKlapdor-Kleingrothaus:2005qv}. There does seem to be some
tension between the derived upper bound from cosmology and this
result. However, at present it seems premature to exclude it based
on cosmological arguments. This is particularly true because the
exact value of the mass inferred from the experiment is highly
uncertain because of the uncertainty in the nuclear matrix element
calculation.

In this section we have discussed only active neutrino species.
However, the framework applies equally well to sterile neutrinos
and other light species. For instance \cite{Dodelson:2005tp} have
derived stringent limits on light sterile neutrinos (see also
\cite{Goswami:2005ng}).

\begin{table}
%\begin{ruledtabular}
\begin{center}
\begin{tabular}{lc}
%\colrule
\hline \hline Case & $m_\nu$ (95\% C.L.) \cr
%\colrule
\hline 1: 11 parameters, CMB, LSS, SNI-a & 2.3 eV \cr 2: 11
parameters, CMB, LSS, SNI-a, BAO & 0.48 eV  \cr 3: 8 parameters,
CMB, LSS, SNI-a, BAO & 0.44 eV \cr 4: 8 parameters, CMB, LSS,
SNI-a, BAO, Ly-$\alpha$ & 0.30 eV \cr \hline Other recent bounds &
\cr \hline 8 parameters, WMAP only \cite{Ichikawa:2004zi} & 2 eV
\cr 8 parameters, WMAP, SDSS \cite{Tegmark:2003ud} & 1.8 eV \cr 8
parameters, WMAP, SDSS, SNI-a, Ly-$\alpha$ \cite{Seljak:2004xh} &
0.42 eV \cr \hline \hline
\end{tabular}
\end{center}
%\end{ruledtabular}
\caption{Mass bounds using different sets of data and parameters
(taken from \cite{newp}.} \label{table:fits}
\end{table}

%%%%%%%%%%%%%%%%%%%%%%%%%%%%%%%%%%%%%%%%%%%%%%%%%%%%%%%%%%%%%%%%
\subsubsection{Neutrino interactions beyond the standard model}
%%%%%%%%%%%%%%%%%%%%%%%%%%%%%%%%%%%%%%%%%%%%%%%%%%%%%%%%%%%%%%%%

The mass bounds derived from cosmology depend on the assumption
that neutrinos have no interactions beyond the standard model weak
interactions. However, experimental constraints on non-standard
neutrino interactions are in general quite weak and allows for the
possibility of new, cosmologically important interactions. One
example is neutrinos coupling to a light scalar or pseudoscalar,
as was discussed in Section \ref{sec:rec}. In this case neutrinos
come into thermal equilibrium with the scalar late in the
evolution of the Universe. This was studied in
\cite{Beacom:2004yd} as a possible way of circumventing the
cosmological bound on neutrino masses. The idea is that neutrinos
coupling to a new massless degree of freedom will annihilate and
disappear once $T \sim m_\nu$, leaving only little imprint on
large scale structure. However, neutrinos which are strongly
coupled will undergo acoustic oscillations prior to CMB formation
instead of damping via free streaming (anisotropic stress). Since
neutrino perturbations act as a source term for the baryon-photon
oscillations this will increase the amplitude of CMB anisotropies
on scales small than the horizon scale at recombination. This
effect was studied numerically in
\cite{Hannestad:2004qu,Trotta:2004ty,Bell:2005dr}. Both studies
find that this effect disfavours strongly coupled neutrinos,
although the exact level of significance differs. The same effect
was also studied analytically in
\cite{Bashinsky:2003tk,Chacko:2003dt,Okui:2004xn}. It should be
noted that \cite{Hannestad:2004qu,Bell:2005dr} studied the effect
only in the strong coupling limit. In the intermediate regime the
situation is more complicated (see \cite{Sawyer:2006ju}).

The fact that strongly interacting neutrinos are disfavoured by
CMB and LSS data means that any model producing neutrino
interactions strong enough to fast momentum exchange around the
epoch of recombination \cite{Hannestad:2005ex}. In the case of
interactions with a light pseudoscalar like the majoron this leads
to a very rough bound on the diagonal coupling $g_{ii}$ of
$10^{-7}$ from consideration of pair creation and annihilation
processes. For the off-diagonal coupling it is roughly $g_{ij} <
10^{-11} (0.05 {\rm eV}/m)^2$ which is in fact the strongest known
bound on these parameters. An exact calculation of the bound would
require a solution of the Boltzmann equation for neutrinos in the
intermediate coupling regime around recombination.

Another possibility which has received quite a lot of attention
recently is the possibility of neutrinos coupling to a very light
scalar field \cite{Fardon:2003eh} (see also
\cite{Peccei:2004sz,Hung:2003jb}). In this case the coupling can
produce an effective neutrino mass which is density dependent
(mass varying neutrinos), and by tuning the coupling it is
possible to have the combined neutrino-scalar fluid have an
effective equation of state close to $w=-1$, i.e.\ dark energy. It
was subsequently argued that the particular form of the potential
in \cite{Fardon:2003eh} leads to a small scale instability which
drives $w$ to 0 \cite{Afshordi:2005ym} (see also
\cite{Takahashi:2006jt}). However, this instability apparently
requires two conditions to be fulfilled: a) The effective mass of
the light scalar should have $m >> H$ and b) The equation of state
should obey $w > -1$. Both these conditions can be violated by
other potentials (see for instance
\cite{Brookfield:2005bz,Kaplinghat:2006jk}). Mass varying
neutrinos have been discussed in other contexts, for instance in
\cite{Barger:2005mn,Cirelli:2005sg,Gonzalez-Garcia:2005xu,Schwetz:2005fy}.

%%%%%%%%%%%%%%%%%%%%%%%%%%%%%%%%%%%%%%%%%%%%%%%%%%%%%%%%%%%%%%%%%%%%%%%%%%%%
\subsubsection{The number of neutrino species - joint CMB and BBN analysis}
%%%%%%%%%%%%%%%%%%%%%%%%%%%%%%%%%%%%%%%%%%%%%%%%%%%%%%%%%%%%%%%%%%%

The BBN bound on the number of neutrino species presented in the
previous section can be complemented by a similar bound from
observations of the CMB and large scale structure. The CMB depends
on $N_\nu$ mainly because of the early Integrated Sachs Wolfe
effect which increases fluctuation power at scales slightly larger
than the first acoustic peak. The large scale structure spectrum
depends on $N_\nu$ because the scale of matter-radiation equality
is changed by varying $N_\nu$.

Many recent papers have used CMB, LSS, and SNI-a data to calculate
bounds bounds on $N_\nu$
\cite{Crotty:2003th,Hannestad:2003xv,Pierpaoli:2003kw,Barger:2003zg,%
Cuoco:2003cu}, and some of the bounds are listed in Table
\ref{table:nnu}. Recent analyses combining BBN, CMB, and large
scale structure data can be found in
\cite{Hannestad:2003xv,Barger:2003zg}, and these results are also
listed in Table \ref{table:nnu}.

Common for all the bounds is that $N_\nu=0$ is ruled out by both
BBN and CMB/LSS. This has the important consequence that the
cosmological neutrino background has been positively detected, not
only during the BBN epoch, but also much later, during structure
formation.

The most recent bound which uses the SNI-a "gold" data set, as
well as the new Boomerang CMB data finds a limit of $N_\nu =
4.2^{+1.7}_{-1.2}$ at 95\% C.L. \cite{Hannestad:2005jj}. The bound
from late-time observations is now as good as that from BBN, and
the two derived value are mutually consistent given the systematic
uncertainties in the primordial helium abundance.

\begin{figure}[htbp]
\begin{center}
\epsfig{file=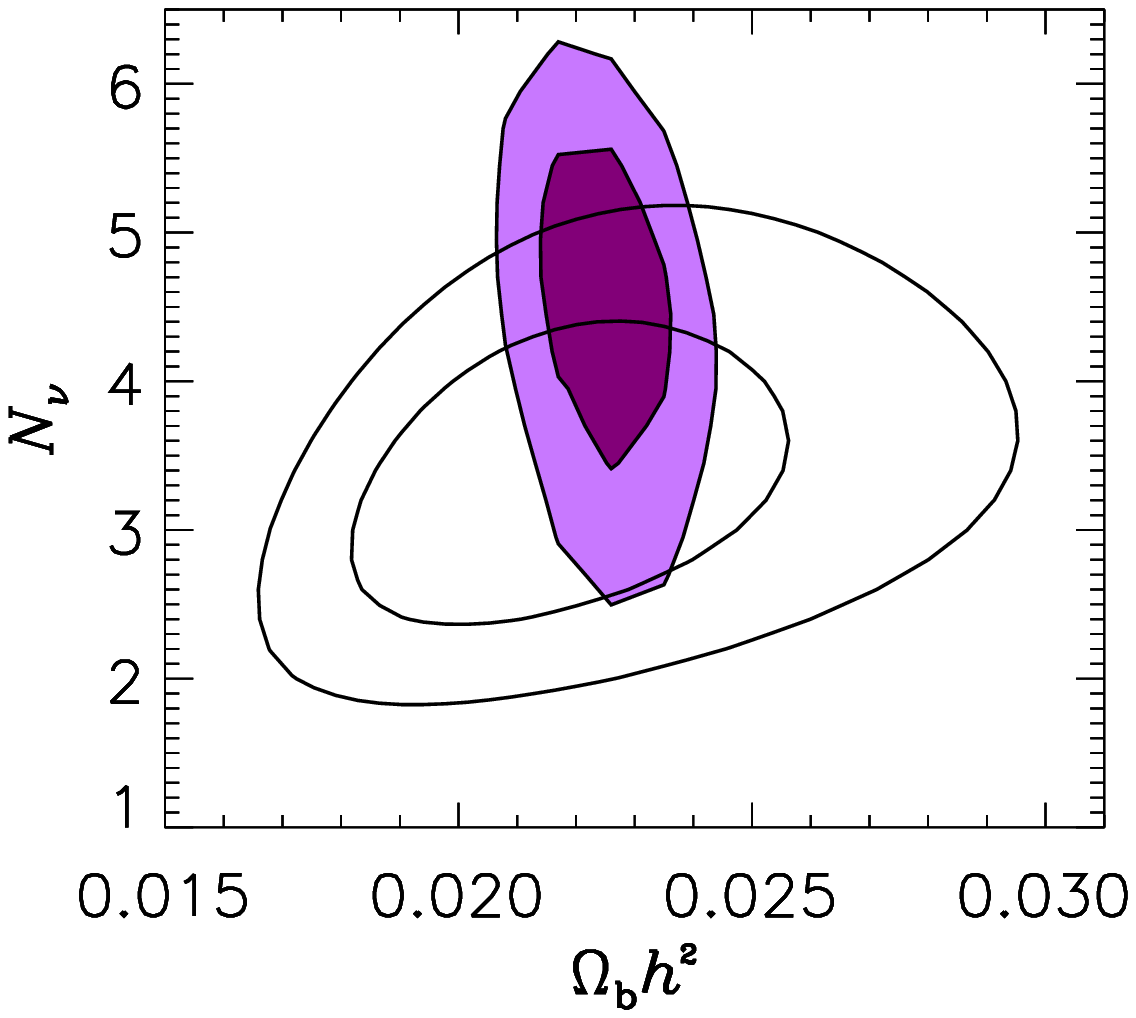,width=0.8\textwidth}
\end{center}
\bigskip
\caption{ The 68\% (dark) and 95\% (light) likelihood contours for
$\Omega_b h^2$ and $N_\nu$ for all available data. The other
contours are 68\% and 95\% regions for BBN, assuming the $^4$He
and D values given in \protect\cite{Cyburt:2004yc}. Figure taken
from \protect\cite{Hannestad:2005jj}.} \label{fig:nnu}
\end{figure}

In Fig.~\ref{fig:nnu} we show the currently allowed region for
$\Omega_b h^2$ and $N_\nu$ (taken from \cite{Hannestad:2005jj}),
showing the overlap between BBN constraints and the CMB+LSS+SNI-a
constraint.

In principle, even if the neutrino mass is very small, it could be
possible to detect the difference between additional relativistic
energy density, parameterized by $N_\nu$, and a non-thermal
distribution of the active neutrinos \cite{Cuoco:2005qr}. However,
it is unlikely that the precision of observational data from
planned experiments will be sufficient to measure the difference.

\begin{table}
\begin{center}
\caption{Various recent limits on the effective number of neutrino
species, as well as the data used.}
\begin{tabular}{@{}lll}
\hline

Ref. & Bound on $N_\nu$ (95\% C.L.) & Data used \\
\hline

Crotty et al. \cite{Crotty:2003th} & $1.4 \leq N_\nu \leq 6.8$ &
CMB, LSS \\

Hannestad \cite{Hannestad:2003xv} & $0.9 \leq N_\nu \leq 7.0$ &
CMB, LSS \\

Pierpaoli \cite{Pierpaoli:2003kw} & $1.9 \leq N_\nu \leq 6.62$ &
CMB, LSS \\

Barger et al. \cite{Barger:2003zg} & $0.9 \leq N_\nu \leq 8.3$ &
CMB \\

Hannestad \cite{Hannestad:2005jj} & $3.0 \leq N_\nu \leq 5.9$ &
CMB, LSS, SNI-a \\

\hline
\end{tabular}
\end{center}
\label{table:nnu}
\end{table}

%%%%%%%%%%%%%%%%%%%%%%%%%%%%%%%%%%%%%%%%%%%%%%%%%%%%%%%%%%%%%%%%%
\subsubsection{Future neutrino mass measurements}
%%%%%%%%%%%%%%%%%%%%%%%%%%%%%%%%%%%%%%%%%%%%%%%%%%%%%%%%%%%%%%%%%

The present bound on the sum of neutrino masses is still much
larger than the mass difference, $|\Delta m_{23}| \sim 0.05$ eV
\cite{Fogli:2003th,Maltoni:2003da}, measured by atmospheric
neutrino observatories and K2K . This means that if the sum of
neutrino masses is anywhere close to saturating the bound then
neutrino masses must the almost degenerate. The question is
whether in the future it will be possible to measure masses which
are of the order $\Delta m_{23}$, i.e.\ whether it can determined
if neutrino masses are hierarchical.

By combining future CMB data from the Planck satellite with a
galaxy survey like the SDSS it has been estimated that neutrino
masses as low as about 0.1 eV can be detected
\cite{Hannestad:2002cn,pastor04} when analyzed within the minimal
8 parameter $\Lambda$CDM model. However, the degeneracy between
$\sum m_\nu$ and $w$ found in present data is likely to persist
even with much more accurate future data \cite{Hannestad:2005gj},
limiting the precision with which the neutrino mass can be
measured.

One very powerful probe which will become available in the future
is weak gravitational lensing. With future CMB experiments like
Planck \cite{planck} and the proposed Inflation Probe
\cite{inflationprobe} it will be possible to measure lensing
effects on the CMB itself. In this case it seems likely that a
sensitivity below 0.1 eV can also be reached with CMB alone
\cite{Abazajian:2002ck,Kaplinghat:2003bh,Lesgourgues:2005yv}.

Another possibility is to probe weak lensing at much lower
redshifts. This can be done by measuring lensing of background
galaxies by large scale structure. This has the major advantage
that it is possible to perform lensing tomography by binning data
in different redshift bins. In \cite{Song:2003gg} the projected
sensitivity from combining future CMB data with a large scale weak
lensing survey was studied. It was estimated that at $1\sigma$ the
sensitivity could be pushed to 0.03 eV. Such lensing surveys are
planned with experiments such as the Large Synaptic Survey
Telescope (LSST) which is planned to start in 2012 \cite{lsst}. At
present the systematic uncertainties related to such surveys is to
some extent unknown. One source of error is related to the
uncertainty in measurements of galaxy shapes and another is
related to the fact the redshift of galaxies must be estimated
photometrically since spectroscopy is not feasible for the large
number of galaxies needed. Estimates in general show that it will
be possible to control the errors at the required level
\cite{Huterer:2005ez,Ma:2005rc}.

Another promising tool which could further increase sensitivity is
to use future cluster surveys \cite{Wang:2005vr}. Such a survey
could also bring the sensitivity to 0.03 eV.

As noted in Ref.~\cite{pastor04} the exact value of the
sensitivity at this level depends both on whether the hierarchy is
normal or inverted, and the exact value of the mass splittings. At
this level, neutrino masses cannot be regarded as degenerate and
the normal and inverted hierarchy models must be tested
separately.

\subsection{Neutrino warm dark matter}

While CDM is defined as consisting of non-interacting particles
which have essentially no free-streaming on any astronomically
relevant scale, and HDM is defined by consisting of particles
which become non-relativistic around matter radiation equality or
later, warm dark matter is an intermediate. One of the simplest
production mechanisms for warm dark matter is active-sterile
neutrino oscillations in the early universe
\cite{Hansen:2001zv,Abazajian:2001vt,Abazajian:2001nj,Shi:1998km,%
Dodelson:1993je}.

One possible benefit of warm dark matter is that it does have some
free-streaming so that structure formation is suppressed on very
small scales. This has been proposed as an explanation for the
apparent discrepancy between observations of galaxies and
numerical CDM structure formation simulations. In general
simulations produce galaxy halos which have very steep inner
density profiles $\rho \propto r^\alpha$, where $\alpha \sim
1-1.5$, and numerous subhalos
\cite{Kazantzidis:2003hb,Ghigna:1999sn}. Neither of these
characteristics are seen in observations and the explanation for
this discrepancy remains an open question. If dark matter is warm
instead of cold, with a free-streaming scale comparable to the
size of a typical galaxy subhalo then the amount of substructure
is suppressed, and possibly the central density profile is also
flattened
\cite{Yoshida:2003rm,Haiman:2001dg,Avila-Reese:2000hg,Bode:2000gq,%
Colin:2000dn,Hannestad:2000gt,jsd} . In both cases the mass of the
dark matter particle should be around 0.5 -- 1 keV
\cite{Hogan:2000bv,Dalcanton:2000hn,Goerdt:2006rw}, assuming that
it is thermally produced in the early universe.

On the other hand, from measurements of the Lyman-$\alpha$ forest
flux power spectrum it has been possible to reconstruct the matter
power spectrum on relatively small scales at high redshift. This
spectrum does not show any evidence for suppression at sub-galaxy
scales and has been used to put a lower bound on the mass of warm
dark matter particles of roughly 500 eV \cite{Viel:2005qj}. An
even more severe problem lies in the fact that star formation
occurs relatively late in warm dark matter models because small
scale structure is suppressed. This may be in conflict with the
low-$l$ CMB temperature-polarization cross correlation measurement
by WMAP which indicates very early reionization and therefore also
early star formation. One recent investigation of this found warm
dark matter to be inconsistent with WMAP for masses as high as 10
keV \cite{Yoshida:2003rm}.

Very interestingly a keV sterile neutrino will decay radiatively
and produce x-ray photons which can reionize the Universe at high
redshift. The decay lifetime is given purely in terms of the mass
and mixing angle with active species
\begin{equation}
\tau = 5 \times 10^{28} \left(\frac{7 \, {\rm keV}}{m}\right)^5
\left(\frac{0.8 \times 10^{-9}}{\sin ^2 \theta}\right)\, {\rm s}.
\end{equation}
Since the total contribution to the energy density is also given
purely in terms of these parameters (assuming a zero lepton
assymetry) it is possible to calculate the total x-ray photon
intensity as a function of redshift.

Such decays have been proposed as a possible explanation for the
very high optical depth indicated by the WMAP polarization
measurement which seems to require partial reionization already at
$z \sim 15-20$
\cite{Hansen:2003yj,Kasuya:2004qk,Mapelli:2005hq,Biermann:2006bu}.
However, in this case the mass must be several keV  in order to
have a short enough lifetime (since $\tau \propto m^{-5}$).
Therefore there does not seem to be a common mass which can
address both the structure formation problems and the reionization
problems.

The case for warm dark matter seems quite marginal, although at
present it is not definitively ruled out by any observations.

%%%%%%%%%%%%%%%%%%%%%%%%%%%%%%%%%%%%%%%%%%%%%%%%%%%%%%%%%%%%%%%%%%%%%%
\section{Discussion} %%%%%%%%%%%%%%%%%%%%%%%%%%%%%%%%%%%%%%%%%%%%%%%%%
%%%%%%%%%%%%%%%%%%%%%%%%%%%%%%%%%%%%%%%%%%%%%%%%%%%%%%%%%%%%%%%%%%%%%%

In the present paper I have discussed how cosmological
observations can be used for probing fundamental properties of
neutrinos which are not easily accessible in lab experiments.
Particularly the measurement of absolute neutrino masses from CMB
and large scale structure data has received significant attention
over the past few years. From cosmological observations it has
been possible to derive an upper bound on the sum of neutrino
masses which seems to be robustly in the 0.5 eV range. This is
substantially better than the present bound from tritium decay
measurements which is $m_{\nu_e} < 2.3$ eV (95\% C.L.), leading to
$\sum m_\nu \lesssim 7$ eV.

In the future this type of measurement will be improved by more
than an order of magnitude by the KATRIN experiment
\cite{katrin,guido} which has a projected sensitivity of 0.2 eV on
the effective electron neutrino mass.

Another cornerstone of neutrino cosmology is the measurement of
the total energy density in non-electromagnetically interacting
particles. For many years Big Bang nucleosynthesis was the only
probe of relativistic energy density, but with the advent of
precision CMB and LSS data it has been possible to complement the
BBN measurement. At present the cosmic neutrino background is seen
in both BBN, CMB and LSS data at high significance, with $N_\nu =
0$ being excluded at more than $5\sigma$.

Finally, cosmology can also be used to probe the possibility of
neutrino warm dark matter, which could be produced by
active-sterile neutrino oscillations.

In the coming years the steady stream of new observational data
will continue, and the cosmological bounds on neutrino will
improve accordingly. For instance, it has been estimated that with
data from the upcoming Planck satellite it could be possible to
measure neutrino masses as low as 0.05 eV which would allow for a
determination of the neutrino mass even in the normal hierarchy.

%%%%%%%%%%%%%%%%%%%%%%%%%%%%%%%%%%%%%%%%%%%%%%%%%%%%%%%%%%%%%%%%%%%%%%
\section*{Acknowledgments} %%%%%%%%%%%%%%%%%%%%%%%%%%%%%%%%%%%%%%%%%%%
%%%%%%%%%%%%%%%%%%%%%%%%%%%%%%%%%%%%%%%%%%%%%%%%%%%%%%%%%%%%%%%%%%%%%%

I acknowledge use of the publicly available CMBFAST package
written by Uros Seljak and Matthias Zaldarriaga~\cite{CMBFAST} and
the use of computing resources at DCSC (Danish Center for
Scientific Computing).

\end{document}